\documentstyle[12pt,amssymb]{article}
\parskip=\medskipamount
   \def\CL{{\cal L}}

   \def\Ga{\Gamma}

\def\la{\lambda}  
   
\def\om{\omega}   
      
\def\IC{{\Bbb C}}
\def\IK{\relax{\rm l\kern-.18 em K}}
\def\IL{\relax{\rm I\kern-.18 em L}}
\def\IN{{\Bbb N}}
\def\IR{{\Bbb  R}}

\def\"i{\'{\i}}
\def\ii{\rm i\,}

\def\Re{\mathop{\rm Re}\nolimits}
\def\Im{\mathop{\rm Im}\nolimits}

\def\Ker{\mathop{\rm Ker}\nolimits}

\def\Id{\mathop{\rm Id}\nolimits}
\def\Image{\mathop{\rm Image}\nolimits}

\def\wh{\widehat}

\def\frac#1#2{{#1\over #2}}

\def\fpd#1#2{\frac{\partial #1}{\partial #2}}
\def\ptos{\leaders\hbox to 2mm{\hfil{.}\hfil}\hfill}
\def\\{\hfill\break}

\font\tenfrak=eufm10  \font\sevenfrak=eufm7  \font\fivefrak=eufm5
\newfam\frakfam
\textfont\frakfam=\tenfrak
\scriptfont\frakfam=\sevenfrak
\scriptscriptfont\frakfam=\fivefrak

\font\tengoth=eufm10 scaled\magstep1 \font\sevengoth=eufm7 \font\fivegoth=eufm5
\newfam\gothfam
\textfont\gothfam=\tengoth\scriptfont\gothfam=\sevengoth
    \scriptscriptfont\gothfam=\fivegoth
\def\goth{\fam\gothfam}    

\newtheorem{proposicion}{Proposition}
\def\today{\ifcase\month\or
     January\or February\or March\or April\or May\or June\or
     July\or August\or September\or October\or November\or December\fi
     \space\number\day, \number\year}

\begin{document}

\title{ Non-symplectic symmetries and bi-Hamiltonian structures of
the rational Harmonic Oscillator }

\author{Jos\'e F. Cari\~nena$^{\dagger}$$^{a,*}$, Giuseppe
Marmo$^{\ddagger}$$^{b}$, and
            Manuel F. Ra\~nada$^{\dagger}$$^{c}$ \\[6pt]
$^{\dagger}$
               {\it Departamento de F\'{\i}sica Te\'orica, Facultad de
Ciencias}\\
               {\it Universidad de Zaragoza, 50009 Zaragoza, Spain}\\[3pt]
$^{\ddagger}$
               {\it Dipartimento  di Scienze Fisiche, Universit\'a
Federico II di Napoli}\\
and {\it INFN, Sezione di Napoli}\\
               {\it Complesso Univ. di Monte Sant'Angelo, Via Cintia, 80125
Napoli, Italy}\\[3pt]
}

\date{}
\maketitle

\begin{quote}
{\bf Abstract.}
The existence of bi-Hamiltonian structures for the rational
Harmonic Oscillator (non-central harmonic oscillator with
rational ratio of frequencies) is analyzed by making use
of the geometric theory of symmetries.
We prove that these additional structures are a
consequence of the existence of dynamical symmetries
of non-symplectic (non-canonical) type.
The associated recursion operators are also obtained.



\end{quote}

\vfill
\footnoterule\noindent
{\small
$^*$ Corresponding author\\
$^{a)}${\it E-mail address:} {jfc@posta.unizar.es} \\
$^{b)}${\it E-mail address:} {giuseppe.marmo@na.infn.it} \\
$^{c)}${\it E-mail address:} {mfran@posta.unizar.es} }
\newpage

\section{Non-symplectic symmetries}

It is well known that there is a close relation \cite{Ma78}
between integrability and the existence of alternatives structures
(see e.g. \cite{Mam02} for a recent paper) and also that integrable
systems are systems endowed with a great number of symmetries.
The purpose of this letter is to analyze, in the particular case of
the $n=2$ harmonic oscillator, how these additional structures arise
from the existence of dynamical symmetries of non-symplectic
(non-canonical) type.

   Let $(M,\om_0,H)$ be a Hamiltonian system and $\Ga_H$ the associated
Hamiltonian vector field, defined by $i(\Ga_H)\,\om_0 = d H$.
A (infinitesimal)  dynamical symmetry of this system is a vector
field $Y\in {\goth X}(M)$ such that $[Y,\Ga_H]=0$.
When  $Y$ is a dynamical but non-symplectic symmetry of the system, 
then we have that (i) the dynamical vector field $\Ga_H$ is bi-Hamiltonian,
 and (ii)
 the function $Y(H)$ is the new Hamiltonian, and therefore it is a constant of
motion.

A sketch of the proof \cite{CaI83}-\cite{Ra00} of this statement is as follows:
The vector field $Y$ does not preserve $\om_0$ and, as it is a
non-canonical transformation, it deter\-mines a new $2$-form
$\om_Y=\CL_{Y}\om_0$ ($\CL_{Y}$ denotes de Lie derivative with respect to $Y$).
As $Y$ is a symmetry, $[Y,\Ga_H]=0$, then
${\cal L}_Y\circ i_{\Gamma_H}=i_{\Gamma_H}\circ {\cal L}_Y$,
and, consequently,
  $$
   i_{\Gamma_H}\,\omega_Y = i_{\Gamma_H}{\cal L}_Y\omega_0
   = {\cal L}_Yi_{\Gamma_H}\omega_0 = {\cal L}_Y(dH)=d(YH)\ .
  $$
Therefore, the 2-form $\om_Y$ is admissible for the dynamical vector field
$\Ga_H$, i.e. ${\cal L}_{\Gamma_H}\omega_Y=0$, which
 is weakly  bi-Hamiltonian with respect to the original
symplectic $2$-form $\om_0$ and the new structure $\om_Y$.
Of course the particular form of $\om_Y$ depends on $Y$ and, in some cases,
it can be just a constant multiple of $\om_0$
(trivial bi-Hamiltonian system).
In some other cases $\om_Y$ may  be a degenerate    $2$-form
with a nontrivial kernel.
In any case, the vector field $\Ga_H$ is a dynamical system solution
of the following two equations
  $$
   i(\Ga_H)\,\om_0 = dH \,,\quad{\rm and}{\quad} i(\Ga_H)\,\om_Y = d[Y(H)]
  $$
Therefore the function $H_Y=Y(H)$, that must be a constant of motion,
can be considered as a new Hamiltonian for $\Ga_H$.

\section{Bi-Hamiltonian structures of the rational Harmonic Oscillator}

  The two-dimensional harmonic oscillator
  \begin{equation}
   H = {1\over 2}\,(p_x^2 + p_y^2)
     + {1\over 2}\,(\la_1^2\, x^2 + \la_2^2\, y^2)\label{Hho2}
  \end{equation}
has the two one-degree of freedom energies, $I_1=E_x$ and $I_2=E_y$,
as fundamental constants of motion.
The superintegrability of the rational case, $\la_1 = m\,{\la_0}$,
$\la_2 = n\,{\la_0}$, with $m,n\in {\IN}$, can be proved by making
use of a complex formalism \cite{{Pe90}, {LoMR99}}.
Let $K_x$, $K_y$, be the following two functions
$K_x = p_x + {\ii} m\,{\la_0}\, x$ and
$K_y = p_y + {\ii} n\,{\la_0}\,y$;
then the Hamiltonian $H$ and the canonical symplectic form
$\omega_0$  become
  $$
   H=\frac 12\left(K_x\, K_x^*+K_y\, K_y^*\right) \,,
  $$
and
  $$
   \omega_0 = \frac{{\ii}}
{2\,m\,\lambda_0}\,{dK_x\wedge dK_x^*}
            + \frac{{\ii}}
{2\,n\,\lambda_0}\,{dK_y\wedge dK^*_y} \,.
  $$
We have
  $$
  \{K_x,K^*_x\} = 2\,{\ii}
\,m\,\lambda_0 \,,{\qquad}
   \{K_y,K^*_y\} = 2\,{\ii}\,n\,\lambda_0 \,,
  $$
and therefore, the evolution equations are
  $$
   \frac{d}{d t}\,K_x   =  \,{\ii} m\,
 {\la_0}\,K_x  \,,{\qquad}
   \frac{d}{d t}\,K_y^* = -\,{\ii} n\,
{\la_0}\,K_y^*\,.
  $$
Hence, the complex function $J$ defined as
  \begin{equation}
   J = K_x^{n}\,(K_y^{*})^{m}  \label{DefJ}
  \end{equation}
is a constant of motion that determines two different real
first integrals, $I_3 = \Im (J)$ and $I_4 = \Re (J)$,
which are polynomials in the momenta of degree $m+n-1$
and $m+n$, respectively.
As an example, for the Isotropic case, $\la_1 = \la_2 = \la_0$, we obtain
\begin{eqnarray}
   \Re(J) &=& p_x\, p_y + {\la_0}^2 \,x\, y  \,,\cr
   \Im(J) &=& {\la_0}\,(x \,p_y - y\, p_x)   \,.\nonumber
\end{eqnarray}
$\Im(J)$ is just the angular momentum, and $\Re(J)$ is the
non-diagonal component of the Fradkin tensor \cite{Fr65}.
For the first non-isotropic case, $\la_1 = \la_0$, $\la_2 = 2\la_0$, 
we arrive to
\begin{eqnarray}
   \Re(J) &=& p_x^2\, p_y + {\la_0}^2\, (4\, y\, p_x - x\, p_y)\,x \,,\cr
   \Im(J) &=& (x \,p_y - y\,p_x) p_x + {\la_0}^2\, x^2 y     \,.\nonumber
\end{eqnarray}
This complex procedure provides not just the fundamental
constant $I_3$, but the pair $(I_3,I_4)$;
although the `partner' function $I_4$ is not independent
(is a function of $I_1$, $I_2$, $I_3$),
we will see that it plays an important r\^ole, since it is
closely concerned with the bi-Hamiltonian formalism.
In fact, we will take the complex function $J$ as our starting point
for the search of symmetries, but  $J$ means not only one
but two functions, $I_3$ and $I_4$.

   The Noether theorem in the Hamiltonian formalism states that
all constants of motion arise from canonical symmetries of the
Hamiltonian function.
Moreover, in differential geometric terms, the infinitesimal
symmetries are simply those corresponding to the Hamiltonian
vector fields, with respect to the canonical structure $\om_0$,
defined by the constants of motion.
In this particular case, the above complex function $J$ given by
(\ref{DefJ}) arises from a symmetry of (\ref{Hho2})
represented by the complex vector field $X_J$ defined by
  \begin{equation}
    i(X_J)\,\om_0 = d J \,,{\quad} X_J(H) = 0 \,.\label{DefXJ}
  \end{equation}
In the following, and for easy of notation, we will suppose ${\la_0}=1$.
\begin{proposicion}
The complex vector field $X_J$, defined by (\ref{DefXJ}) as the
canonical infinitesimal symmetry associated to $J$, can be written
as a linear combination of two dynamical but non-symplectic symmetries
of $\Gamma_H$.
\end{proposicion}
{\sl Proof:} Let us denote by $Y_{xm}$ and  $Y_{yn}$ the Hamiltonian
vector fields of $K_x$ and  $K_y$
  $$
   i(Y_{xm})\,\om_0 = d K_x \,,{\qquad}
   i(Y_{yn})\,\om_0 = d K_y \,,
  $$
with coordinate expressions
  $$
   Y_{xm} = \fpd{}{x} - {\ii}m\fpd{}{p_x} \,,{\qquad}
   Y_{yn} = \fpd{}{y} - {\ii}n\fpd{}{p_y} \,.\quad
  $$
Notice that, as $H=I_1+I_2$ with $|\,K_x\,|^2 = 2I_1$ and
$|\,K_y\,|^2 = 2I_2$, we have
  $$
   \Gamma_H = \Re\,(K_x^*\,Y_{xm} + K_y\,Y_{yn}^*) \,.
  $$
Then, the complex vector field $X_J$, canonical infinitesimal symmetry
of the Harmonic oscillator, can be written as the following linear
combination
  $$
   X_J = n\,Y + m\,Y' \,,
  $$
where the $Y$, $Y'$, are given by
  $$
   Y = (K_x^{(n-1)}K_y^{*m})\,Y_{xm}   \,,{\quad}
   Y'= (K_x^{n}K_y^{*(m-1)})\,Y_{yn}^* \,.
  $$
The important point is that these two vector fields, $Y$ and $Y'$,
are neither locally-Hamiltonian with respect to $\omega_0$
  $$
   {\cal L}_Y\omega_0\ne 0  \,,{\qquad}  {\cal L}_{Y'}\omega_0\ne 0 \,,
  $$
nor infinitesimal symmetries of the Hamiltonian
  $$
   {\cal L}_YH\ne 0  \,,{\qquad} {\cal L}_{Y'}H\ne 0 \,.
  $$
Concerning the Lie bracket of $Y$ with the dynamical vector field
$\Gamma_H$, it is given by
  $$
   [Y,\Gamma_H] = (K_x^{n-1}K_y^{*m}) [Y_{xm},\Gamma_H]
                - \Gamma_H(K_x^{n-1}K_y^{*m})\,Y_{xm}
  $$
but as
  $$
   [Y_{xm},\Gamma_H] = - {\ii} Y_{xm}  \,,{\qquad}
   [Y_{yn}^*,\Gamma_H] = {\ii} Y_{yn}^*\,,
  $$
and
  \begin{eqnarray}
    \Gamma_H(K_x^{n-1}K_y^{*m})
     &=& (n-1)({\ii}m)(K_x^{n-1}K_y^{*m}) + m(-{\ii}n)(K_x^{n-1}K_y^{*m}) \cr
     &=& -{\ii}m\,(K_x^{n-1}K_y^{*m}) \,,             {\nonumber}
  \end{eqnarray}
we arrive to
  $$
   [Y,\Gamma_H] = 0 \,.
  $$
Thus $Y$ is a dynamical but non-symplectic (non-canonical) symmetry
of $\Gamma_H$.
It can be proved, in a similar way, that this property is also true
for $Y'$.
Notice that, in the language of 1-forms, this property arises from the
fact that $dJ$ splits as a sum of two non-closed 1-forms that, nevertheless,
remain invariant under $\Gamma_H$, that is,
$dJ = n\,\phi_1 + m\,\phi_2$,
$d\phi_r \ne 0$, ${\cal L}_{\Gamma_H}(\phi_r)=0$, $r=1,2$.

   Two new structures can be obtained from $\omega_0$ by Lie derivation
with respect to $Y$ and $Y'$.
If we denote by $\omega_Y$ and $\omega'_Y$ these two new 2-forms,
$\omega_Y={\cal L}_Y\omega_0$ and $\omega'_Y={\cal L}_{Y'}\omega_0$,
then we obtain
  $$
   \omega_Y = -m\,(K_x^{(n-1)}K_y^{*(m-1)})\,dK_x{\wedge}dK_y^* \,,{\quad}
   \omega'_Y=  n\,(K_x^{(n-1)}K_y^{*(m-1)})\,dK_x{\wedge}dK_y^* \,.
  $$
In the following we will denote by $\Omega$ the complex 2-form defined as
  $$
   \Omega = dK_x{\wedge}dK_y^* = \Omega_1 + {\ii}\Omega_2
  $$
where the two real 2-forms, $\Omega_1=\Re(\Omega)$ and
$\Omega_2=\Im(\Omega)$, take the form
  $$
   \Omega_1 = m\,n\,dx{\wedge}dy +dp_x{\wedge}dp_y  \,,{\qquad}
   \Omega_2 = m\,dx{\wedge}dp_y  +n\,dy{\wedge}dp_x \,.
  $$

   Notice that $\omega_Y$ and
 $\omega'_Y$ satisfy the relation
$n\,\omega_Y+m\,\omega'_Y=0$.
Actually, this can be considered as a consequence of the fact that
$X_J$ is locally Hamiltonian with respect to the canonical form
$\omega_0$.

\begin{proposicion}
The dynamical vector field $\Gamma_H$ of the rational Harmonic Oscillator
is a bi-Hamiltonian system with respect to $(\omega_0,\omega_Y)$.
\end{proposicion}
{\sl Proof:} Notice that
  $$
   i(\Gamma_H)\omega_Y =
     -m\,(K_x^{(n-1)}K_y^{*(m-1)})\,i(\Gamma_H)\Omega\ ,
  $$
and as
  $$
   i(\Gamma_H)\Omega = \Gamma_H(K_x)dK_y^* - \Gamma_H(K_y^*)dK_x
                     = {\ii}m\,K_x\,dK_y^* + {\ii}n\,K_y^*\,dK_x \,,
  $$
we obtain that
  $$
   i(\Gamma_H)\omega_Y = -{\ii}m\,d(K_x^n\,K_y^{*m}) \,.
  $$
Thus, $\Gamma_H$ is Hamiltonian vector field with respect to $\omega_Y$
with 
$K_x^n\,K_y^{*m}$ as Hamiltonian function.
Moreover, we can also compute the action of $Y$ on $H$;
a direct calculation gives
  $$
   H_Y {\equiv} Y(H) = - {\ii}m\,(K_x^n\,K_y^{*m})\ .
  $$
To conclude, we have found that the integral of motion $J$
determines the following bi-Hamiltonian system
  $$
   i(\Gamma_H)\omega_0 = dH \,,{\quad}
   i(\Gamma_H)\omega_Y = d\,H_Y \,.
  $$

Remark first that $\Gamma_H$ is bi-Hamiltonian with respect to two different
structures: the canonical symplectic form $\omega_0$ and another one,
$\omega_Y$, which is complex.
If we write $\omega_Y=\omega_4+{\ii}\,\omega_3$, then $\Gamma_H$
can be considered as a bi-Hamiltonian system with respect to the
following three real forms $(\omega_0,\omega_3,\omega_4)$
(i.e. it is  a three-Hamiltonian system).
The $\omega_0$-Hamilton equation determined by $J$,
  $$
   i(X_J)\,\omega_0 = dJ\ ,
  $$
is also complex; thus it determines two real Hamiltonian equations
  $$
   i(X_4)\,\omega_0 = dI_4 \,,{\qquad}
   i(X_3)\,\omega_0 = dI_3 \,,
  $$
with $X_4$, $X_3$, given by $X_J=X_4+{\ii}\,X_3$.

As a second remark, the complex 2-form $\Omega = dK_x{\wedge}dK_y^*$
is well defined but it is not symplectic.
In fact, it can be proved that $\Omega_1=\Re(\Omega)$ and
$\Omega_2=\Im(\Omega)$ satisfy
  $$
   \Omega_1{\wedge}\Omega_1 = \Omega_2{\wedge}\Omega_2
         = m n\,(dx{\wedge}dy{\wedge}dp_x{\wedge}dp_y)
        \,,{\qquad}{\rm and}{\qquad}
   \Omega_1{\wedge}\Omega_2 = 0\ ,
  $$
so we obtain
  $$
   \Omega{\wedge}\Omega =
   (\Omega_1{\wedge}\Omega_1 - \Omega_2{\wedge}\Omega_2)
         + 2\,{\ii} \Omega_1{\wedge}\Omega_2 = 0 \,.
  $$
Thus, the degenerate character of $\Omega$ is directly related
with its complex nature.  Moreover, the kernel of $\Omega$ is
the distribution generated by $Y_{xm}$ and $Y_{yn}^*$,
  $$
   \Ker\Omega = \{\, f\,Y_{xm} + g\,Y_{yn}^* \mid
         f,g\,:\,\IR^2{\times}\IR^2\to{\IC}\,\} \,,
  $$
therefore it satisfies
  $$
   [\,\Ker\Omega\,,\,\Gamma_H\,] \ \subset\ \Ker{\Omega} \,.
  $$

    Finally, the 2-form $\omega_Y$ is also degenerate.
We obtain, $\Ker\omega_Y=\Ker\Omega$, because of the
relation between $\Omega$ and $\omega_Y$.
However $\omega_3$ and $\omega_4$, defined as
$\omega_Y=\omega_4+{\ii}\omega_3$, are symplectic real forms.
Moreover,  the  form $\omega_0+\Omega$ is symplectic because
of $\{K_x,K^*_y\}=0$.

\section{Recursion operators }

  The bi-Hamiltonian structure $(\omega_0,\omega_Y)$ defines a
complex recursion operator $R_Y$ by
  $$
   \omega_Y(X,Y)=\omega_0(R_YX,Y) \,,{\quad}
   \forall X,Y\in {\goth X}(M) \,,
  $$
or, equivalently, $R_Y = \wh{\omega}_0^{-1}{\circ}\wh{\omega}_Y$.
Since it is complex, it can be written as $R_Y=R_4+{\ii}R_3$, so that
$R_4$ and $R_3$ satisfy the relations
  $$
   \omega_3(X,Y)=\omega_0(R_3X,Y) \,, {\quad}{\rm and}{\quad}
   \omega_4(X,Y)=\omega_0(R_4X,Y) \,.
  $$
Thus, we have that $R_3$ and $R_4$ are given by
$R_3=\wh{\omega}_0^{-1}{\circ}\wh{\omega}_3$ and
$R_4=\wh{\omega}_0^{-1}{\circ}\wh{\omega}_4$.

    The important point is that the complex 2-form
$\Omega=dK_x{\wedge}dK_y^*$ can be decomposed as
$\Omega = \Omega_1+{\ii}\Omega_2$, where both 2-forms,
$\Omega_1$ and $\Omega_2$, are symplectic.
Hence, we have, in addition to $R_3$ and $R_4$, two other recursion
operators $R_1$ and $R_2$ associated with the bi-Hamiltonian structures
provided by $\Omega_1$ and $\Omega_2$, respectively.

\begin{proposicion}
The tensor fields $R_1$ and $R_2$ are invertible operators which
anticommute and satisfy $R_2^2=R_1^2$.
\end{proposicion}
{\sl Proof:} As $\Omega_1$ and $\Omega_2$ are symplectic forms,
the operators $R_1$ and $R_2$ are invertible.
Their coordinate expressions are given by
   \begin{eqnarray}
    R_1 &=& \fpd{}{y} \otimes dp_x - \fpd{}{x}\otimes dp_y +
     m\,n\, \left(\fpd{}{p_x}\otimes dy - \fpd{}{p_y}\otimes dx \right)  \,,\\
    R_2 &=& m\left(\fpd{}{y}\otimes dx + \fpd{}{p_x}\otimes dp_y\right)
          + n\left(\fpd{}{x}\otimes dy + \fpd{}{p_y}\otimes dp_x \right)\,.
   \end{eqnarray}
Therefore,
  $$
   R_1^2 = m\,n\,{\Id}  \,,{\qquad} R_2^2 = m\,n\,{\Id} \,.
  $$
Moreover we have
  $$
   R_2\,R_1 = n\left(\fpd{}{x} \otimes dp_x - m^2\fpd{}{p_x} \otimes dx\right)
            - m\left(\fpd{}{y} \otimes dp_y - n^2\fpd{}{p_y} \otimes 
dy\right) \,,
  $$
and $R_1R_2 = -\,R_2R_1$; therefore $R_2R_1+R_1R_2=0$.
{\smallskip}

We recall that the relation between $\omega_Y$ and $\Omega$ is
$\omega_Y=\omega_4+{\ii}\,\omega_3 = -m\,{\IK}\,\Omega$,
with the complex function ${\IK}$ given by
${\IK}=K_x^{(n-1)}K_y^{*(m-1)} = {\IK}_r+{\ii}\, {\IK}_i$.
Thus, the above two tensor fields, $R_3$ and $R_4$, are given by
  \begin{eqnarray}
    R_4 &=& -m\,\left({\IK}_rR_1 - {\IK}_iR_2\right) \,, \\
    R_3 &=& -m\,\left({\IK}_rR_2 + {\IK}_iR_1\right) \,,
  \end{eqnarray}
and, making use of the preceding proposition, we arrive to
  $$
   R^2_4 = R_3^2=r\,\Id  \,,{\quad}{\rm with}{\quad}
   r=m^2(mn)^2\,|{\IK}|^2  \,,
  $$
and
  $$
   R_4R_3 = m^2\,|{\IK}|^2\,R_1R_2 \,,{\quad}
   R_3R_4 = m^2\,|{\IK}|^2\,R_2R_1 \,,
  $$
where the modulus of ${\IK}$ is function of the two first integrals,
$I_1$ and $I_2$,
  $$
   |{\IK}|^2 = {\IK}_r^2+{\IK}_i^2=(2I_1)^{(n-1)}\,(2I_2)^{(m-1)} \,.
  $$
Thus, the two tensor fields, $R_3$ and $R_4$, anticommute as well.

\begin{proposicion}
  The complex operator $R_Y=R_4+{\ii}R_3$ is such that
  $\Image(R_Y)=\Ker R_Y$.
\end{proposicion}
{\sl Proof:} Let $X$ be a generic vector field  on the phase space
  $$
   X = a\,\fpd{}{x} + b\,\fpd{}{y} + c\,\fpd{}{p_x} + d\,\fpd{}{p_y}\ ,
  $$
with $a,b,c,d$, arbitrary functions. Then, since $R_Y=R_4+{\ii}R_3$
is given by $R_Y= -m\,{\IK}\,(R_1+{\ii}\,R_2)$, the subspace Image
of $R_Y$ is given by
  $$
   \Image(R_Y) = - \Bigl\{\,Z\in{\goth X}(\IR^2{\times}\IR^2)\ |\
             Z = - m\,{\IK}\,\bigl( R_1(X)+{\ii}R_2(X)\bigr) \,\Bigr\} \,.
  $$
We obtain
  $$
   R_1(X)+{\ii}R_2(X) = - (d-{\ii}\,n\,b)Y_{xm}+(c+{\ii}\,m\,a)Y_{yn}^* \,.
  $$
Thus, $\Image(R_Y)$ is made up of linear combinations of $Y_{xm}$
and $Y_{yn}^*$ with arbitrary complex functions as coefficients.
But, since  $\Ker R_Y = \Ker\omega_Y$ and $\Ker\omega_Y$ coincides
with $\Ker\Omega$, which is also spanned by $Y_{xm}$ and $Y_{yn}^*$,
we arrive to $\Image(R_Y)=\Ker R_Y$.
Consequently $R_Y^2 = R_Y{\circ}R_Y =0$.
{\smallskip}

   Given a bi-Hamiltonian system on a manifold $M$,
$i(\Gamma)\,\omega_0 = dH_0 $ and $i(\Gamma)\,\omega_1 = dH_1$,
the point is that the tensor field $R$, that was just defined by
the relation between $\omega_1$ and $\omega_0$, induces a sequence
of structures.
Starting with the basic Hamiltonian system $(\omega_0,\Ga_0=\Gamma,dH_0)$
we can construct a sequence of 2-forms $\omega_k$, of vector fields
$\Gamma_k$, and of 1-forms $\alpha_k$, $k=1,2,\dots$, defined by
$\wh\omega_k=\wh\omega_0\circ R^k$,
$\Gamma_k=R^k(\Gamma_0)$, and
$\alpha_k=R^k(dH_0)$.
Then it follows that
  \begin{eqnarray}
    &&i(\Gamma_0)\,\omega_1 = i(\Gamma_1)\,\omega_0 = dH_1 \,,\cr
    &&i(\Gamma_0)\,\omega_2 = i(\Gamma_1)\,\omega_1 =
      i(\Gamma_2)\,\omega_0=\alpha_2 \,,{\nonumber}
  \end{eqnarray}
where
  \begin{eqnarray}
      \wh\omega_1 = \wh\omega_0{\circ}R  \,,{\qquad}
    &&\wh\omega_2 = \wh\omega_1{\circ}R  \,, \cr
      \Gamma_1 = R(\Gamma_0)\,,{\qquad}
    &&\Gamma_2 = R(\Gamma_1)\,,          \cr
       dH_1 = R^*(dH_0)    \,,{\qquad}
    &&\alpha_2 = R^*(dH_1)\,,        {\nonumber}
  \end{eqnarray}
The 1-form $\alpha_2$ is not necessarily exact, but if there is
$H_2$ such that $\alpha_2=dH_2$, then the vector field $\Gamma_1$
is a bi-Hamiltonian system as well.
An interesting case is when $\alpha_2$ is not exact but there exist
a nonvanishing function $F_2$ and another function 
$H_2$ such that $\alpha_2 = F_2\,dH_2$.
Then  $F_2^{-1}$ is an integrating factor for $\alpha_2$,
and the vector field $\Gamma_1$ is said to be quasi-bi-Hamiltonian
\cite{BrC96},\cite{MoT97}.

   Coming back to the rational harmonic oscillator as a bi-Hamiltonian
system, $i(\Gamma_H)\,\omega_0=dH_0$ and $i(\Gamma_H)\,\omega_Y=dH_Y$,
the situation is as follows:

(i) The action of $R_Y$ is such that $\Gamma_H{\equiv}\Gamma_0$ becomes
$\Gamma_1=R_Y(\Gamma_H) = -\,{\ii}m\,X_J$,

(ii) $dH_0$ transforms into $dH_Y=R_Y^*(dH_0) = -\,{\ii}dJ$, and

(iii) $\omega_0$ becomes $\omega_Y$ such that $\wh{\omega}_Y =
\wh{\omega}_0{\circ}R_Y$.

We have proved that $R_Y^2=0$ because of Proposition 4;
therefore, $\Gamma_1$ transforms into the new field,
$\Gamma_2=R_Y(\Gamma_1)=R_Y^2(\Gamma_H)=0$,
while $dH_Y$ transforms into $\alpha_2=R_Y^*(dH_Y)=R_Y^{*2}(dH_0)=0$.
Hence, it follows that the equation
  $$
   i(\Gamma_H)\,\omega_2 = i(\Gamma_1)\,\omega_1 =
   i(\Gamma_2)\,\omega_0 = \alpha_2
  $$
becomes
  $$
   i(\Gamma_1)\,\omega_1=0 \,.
  $$
Notice that this last equation corresponds to the property
$i(X_J)\omega_Y=0$.

   The Harmonic Oscillator can be considered as a complex and weakly
bi-Hamiltonian system, or alternatively, as endowed with two different
real bi-Hamiltonian structures
  $$
   i(\Gamma_H)\,\omega_0 = dH_0    \,,{\quad}
   i(\Gamma_H)\,\omega_4 = m\,dI_3 \,,{\quad}
   i(\Gamma_H)\,\omega_3 = -m\,dI_4\,.
  $$
One real structure gives rise to $\Gamma_{13}$ defined by
$\Gamma_{13} = R_3(\Gamma_H)$, and the other one to
$\Gamma_{14} = R_4(\Gamma_H)$. They are such that
  $$
   i(\Gamma_{14})\,\omega_0 = i(\Gamma_H)\,\omega_4 =  m\ dI_3 \,,{\quad}
   i(\Gamma_{13})\,\omega_0 = i(\Gamma_H)\,\omega_3 = -m\ dI_4 \,.
  $$
Moreover taking into account that $R_4^2 = r\,\Id$, we arrive to
  \begin{eqnarray}
    \Gamma_{24} &=& R_4(\Gamma_{14}) = R_4^2(\Gamma_H) = r\,\Gamma_H \,,\cr
    \alpha_{24} &=& R_4^*(m\,dI_3) = R_4^{*2}(dH_0) = r\,dH_0        \,,\cr
    \wh\omega_{24} &=& \wh\omega_4{\circ}R_4 = \wh\omega_0{\circ}R_4^2
                      = r\,\wh\omega_0\,,                {\nonumber}
  \end{eqnarray}
and similar results for $R_3$.

  Now, making use of all these relations, we can prove the following
final proposition concerning the properties of the vector fields
$X_3$ and $X_4$.
\begin{proposicion}
Let $X_3$ and $X_4$ denote the two infinitesimal canonical
symmetries generating the two constants of motion $I_3$ and $I_4$.
Then $X_3$ and $X_4$ are quasi-bi-Hamiltonian systems.
Moreover, $\omega_4(X_3,\Gamma_H)=\omega_3(X_4,\Gamma_H)=0$.
\end{proposicion}
{\sl Proof:} The rational Harmonic Oscillator is endowed with the two
constants of motion $I_3$ and $I_4$ which means, via the Hamiltonian
Noether theorem, the existence of two symmetries.
They are geometrically represented by two vector fields, $X_3$ and $X_4$,
that can be uniquely determined as solutions of the following two equations
  $$
   i(X_3)\,\omega_0=dI_3\,,{\qquad}  i(X_4)\,\omega_0=dI_4\,.
  $$
Then we have
  $$
   \Gamma_{13}=R_3(\Gamma_H) = -m\,X_4  \,,{\qquad}
   \Gamma_{14}=R_4(\Gamma_H) =  m\,X_3  \,.
  $$
Hence, if we denote by $f_{34}$ the function
  $
   f_{34} = m (mn)^2\, |{\IK}|^2 \,,
  $
we arrive to
  $$
   i(X_3)\,\omega_0= dI_3 \,,{\qquad}
   i(X_3)\,\omega_4= f_{34}\,dH_0 \,,
  $$
and
  $$
   i(X_4)\,\omega_0= dI_4 \,,{\qquad}
   i(X_4)\,\omega_3=-f_{34}\,dH_0 \,.
  $$
So, both $X_3$ and $X_4$ are quasi-bi-Hamiltonian systems.

A direct consequence of this property is that the dynamical vector field
$\Gamma_H$ is orthogonal to $X_3$ with respect to the symplectic structure
$\omega_4$,
  $$
   i(X_3)i(\Gamma_H)\,\omega_0=0  \,,{\quad}{\rm and}{\quad}
   i(X_3)i(\Gamma_H)\,\omega_4=0  \,.
  $$
Similarly, we obtain
  $$
   i(X_4)i(\Gamma_H)\,\omega_0=0  \,,{\quad}{\rm and}{\quad}
   i(X_4)i(\Gamma_H)\,\omega_3=0  \,.
  $$
Finally, $X_3$ and $X_4$ are orthogonal vector fields with respect to
both structures, $\omega_3$ and $\omega_4$:
  $$
   i(X_3)i(X_4)\,\omega_3=0  \,,{\quad}{\rm and}{\quad}
   i(X_3)i(X_4)\,\omega_4=0  \,.
  $$

{\small
\section*{\bf Acknowledgements.}
This work is dedicated to Prof. Domingo Gonz\'alez, from the University of
Zaragoza, on the occasion of his retirement.
Support of Spanish DGI projects, BFM-2000-1066-C03-01 and FPA-2000-1252,
is acknowledged.


\end{document}